\documentclass[a4paper,10pt,conference]{IEEEtran}

\usepackage{latexsym}
\usepackage{graphicx}
\usepackage[mathscr]{eucal}
\usepackage{cite}
\usepackage{subcaption}
\usepackage{comment}
\usepackage{amsthm}
\usepackage{overpic}
\usepackage{steinmetz}
\usepackage{array}
\usepackage{url}
\usepackage{xcolor}
\IEEEoverridecommandlockouts

\theoremstyle{plain}

\newtheorem{remark}{Remark}

\usepackage{latexsym}
\usepackage{graphicx}
\usepackage{amsfonts,amssymb,amsmath}
\usepackage{varwidth}
\usepackage[T1]{fontenc}
\usepackage{cite}
\usepackage{subcaption}
\usepackage{comment}
\usepackage{overpic}
\usepackage{steinmetz}
\usepackage{array}
\usepackage{url}
\usepackage{xcolor}
\usepackage{algorithm,algorithmic}

\IEEEoverridecommandlockouts

\usepackage{fancyhdr}
\begin{document}

\title{\textcolor{black}{Channel-Coherence-Adaptive Two-Stage Fully Digital Combining for mmWave MIMO Systems}
\thanks{This work was supported by the Smarter Electronics System program by Vinnova and by the Wallenberg Academy Fellow program.}
}

\author{\IEEEauthorblockN{Yasaman Khorsandmanesh$^*$, Emil Björnson$^*$, Joakim Jaldén$^*$, and Bengt Lindoff\IEEEauthorrefmark{3}}
\IEEEauthorblockA{
\textit{$^*$KTH Royal Institute of Technology, Stockholm, Sweden \IEEEauthorrefmark{3}BeammWave AB, Lund, Sweden}\\
Email: \{yasamank, emilbjo, jalden\}@kth.se,  bengt@beammwave.com}}

\maketitle

\begin{abstract}
This paper considers a millimeter-wave wideband point-to-point MIMO system with fully digital transceivers at the base station and the user equipment (UE), focusing on mobile UE scenarios. A main challenge when building a digital UE combining is the large volume of baseband samples to handle. To mitigate computational and hardware complexity, we propose a novel two-stage digital combining scheme at the UE. The first stage reduces the $N_{\text{r}}$	received signals to $N_{\text{c}}$ streams before baseband processing, leveraging channel geometry for dimension reduction and updating at the beam coherence time, which is longer than the channel coherence time of the small-scale fading. By contrast, the second-stage combining is updated per fading realization.
We develop a pilot-based channel estimation framework for this hardware setup based on maximum likelihoodestimation in both uplink and downlink.
Digital precoding and combining designs are proposed, and a spectral efficiency expression that incorporates imperfect channel knowledge is derived. The numerical results demonstrate that the proposed approach outperforms hybrid beamforming, showcasing the attractiveness of using two-stage fully digital transceivers in future systems.

\end{abstract}
\begin{IEEEkeywords}
Wideband mmWave point-to-point MIMO, Mobile UE, Two-Stage Digital Combining, Channel Estimation.
\end{IEEEkeywords}
\section{Introduction}
Millimeter-wave (mmWave) technology holds great potential for wireless networks due to its significantly larger bandwidth compared to sub-6 GHz systems \cite{rangan2014millimeter}. However, mmWave systems have faced deployment challenges in 5G. Key limitations include poor penetration through obstacles like walls, which limits the coverage in urban environments, and environmental factors like rain and fog that heavily attenuate mmWave signals, complicating service reliability. 
Furthermore, effective beamforming for mmWave currently requires complex, power-intensive hardware, leading to rapid battery drain at the user equipment (UE) \cite{dutta2019case}. These limitations suggest that the widespread mmWave deployment will wait until significant advancements in hardware design, power management, and signal processing have occurred.  

Previous research emphasized reducing the number of radio frequency (RF) chains in mmWave systems, as each requires costly and power-intensive components like mixers, filters, and analog-to-digital converters (ADCs).
{\color{black}The literature explores various hardware architectures, including analog (ABF), hybrid (HBF), and fully digital beamforming (DBF) \cite{sohrabi2017hybrid,yu2016alternating,tsai2018sub}. ABF and HBF in mmWave systems for UEs pose significant design challenges, including reliance on resource-inefficient codebooks,  beam sweeping, and limitation to LoS channels.} Analog phase shifters for beam steering cause significant heat dissipation issues, leading to complicated thermal management and reduced output power. 

For these reasons, there is a renewed interest in DBF since it offers greater flexibility for beam control, multi-user support, and responsiveness to rapidly changing mmWave channels \cite{lindoff2021ultimate}.DBF and HBF require a similar number of RF components \cite[Ch.~7]{bjornson2024introduction}, thus, the basic premise for the HBF is that phase shifters have lower power and cost than ADCs \cite{beammwavewhite}.
However, if DBF is implemented with a lower ADC resolution, it can provide better energy efficiency \cite{roth2018comparison}.

This paper considers a new DBF approach to single-user multiple-input multiple-output (MIMO), where the UE has reduced digital hardware capabilities.
It can sample the received downlink signal at each of its $ N_{\mathrm{r}}$ antennas but must apply a two-stage combining scheme. {\color{black} 
At the first stage, the signal dimension is reduced to $ N_{\mathrm{c}}$, consequently decreasing the signaling and processing demands on the baseband processor. In the second stage, digital combining can be conducted with matrix dimensions matching those in a conventional HBF approach. 
Using the beam coherence concept \cite{khorsandmanesh2024beam}, we show that the first-stage combining can be updated infrequently while only the second-stage combining must track the small-scale fading variations.
The proposed method is practically feasible and could be implemented on the BeammWave technology platform \cite{lindoff2021ultimate}.
We develop the necessary signal processing algorithms for this novel DBF setup and quantify the communication performance under UE mobility.
The main contributions are:

\begin{itemize}
    \item We propose a practical two-stage digital combining scheme for mmWave devices, where the stages are updated at different time granularities, based on the beam coherence concept from \cite{khorsandmanesh2024beam}.

    \item We develop a pilot-based maximum likelihood (ML) channel estimation framework to set the digital precoding and two stages of combining matrices.
    \item We evaluate the spectral efficiency (SE) with imperfect channel state information (CSI) by deriving a novel SE expression and propose precoding and combining schemes suitable for this setup.
\end{itemize}
Numerical results confirm that the two-stage digital combining greatly enhances SE compared to HBF and performs close to the upper bound without hardware limitations.}
\begin{figure}[!t]
    \centering 
\includegraphics[width=0.4\columnwidth]{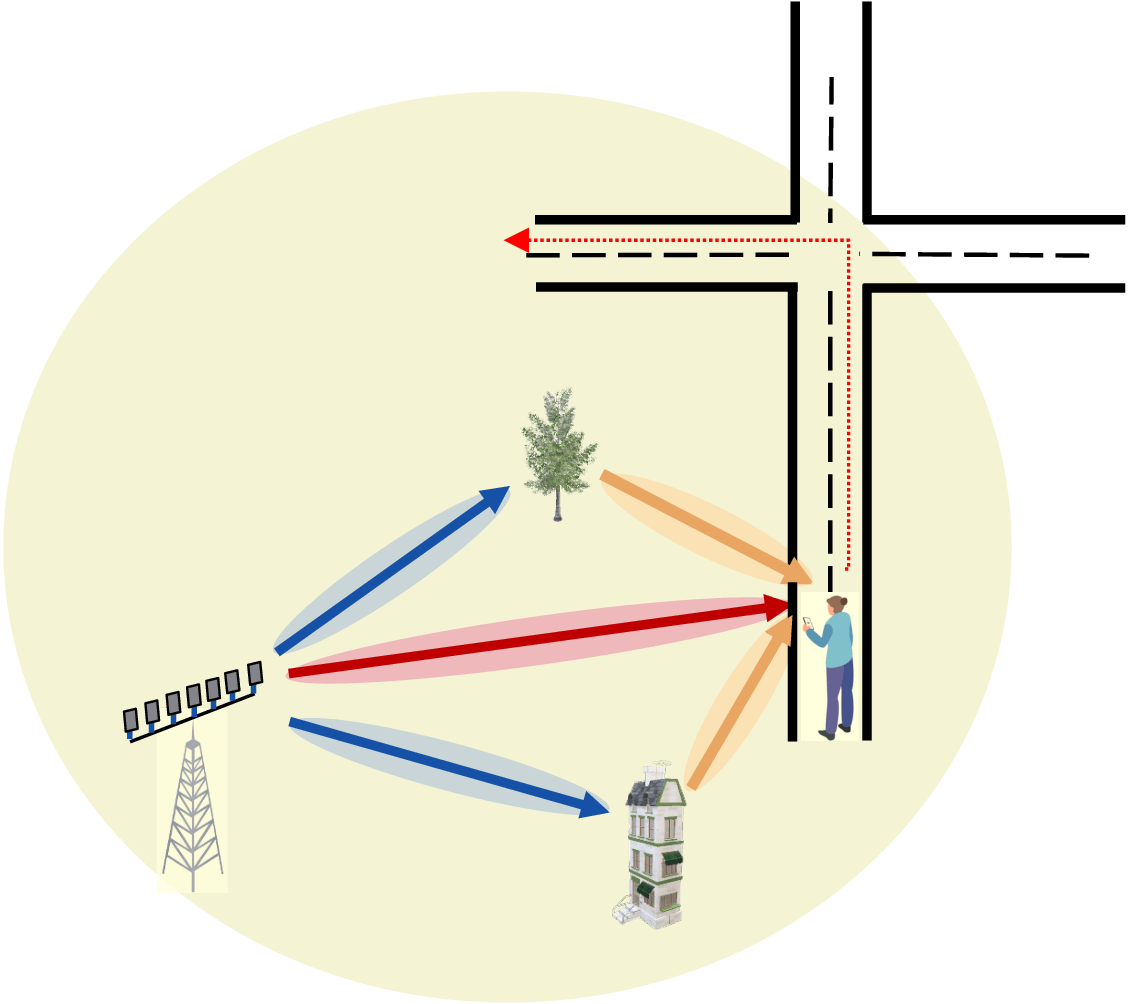}   \caption{ {\color{black}  A mmWave point-to-point MIMO mobile system.}} 
\label{fig:systemmodel}
\end{figure}
\section{System Model} \label{sec:sys}
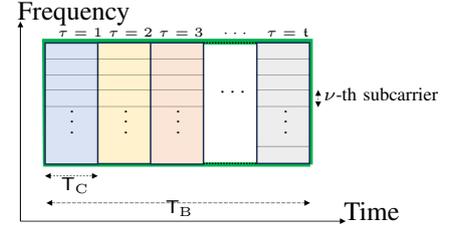
\begin{figure}[t!]
  \centering
   \begin{overpic}
    [width=0.5\columnwidth]{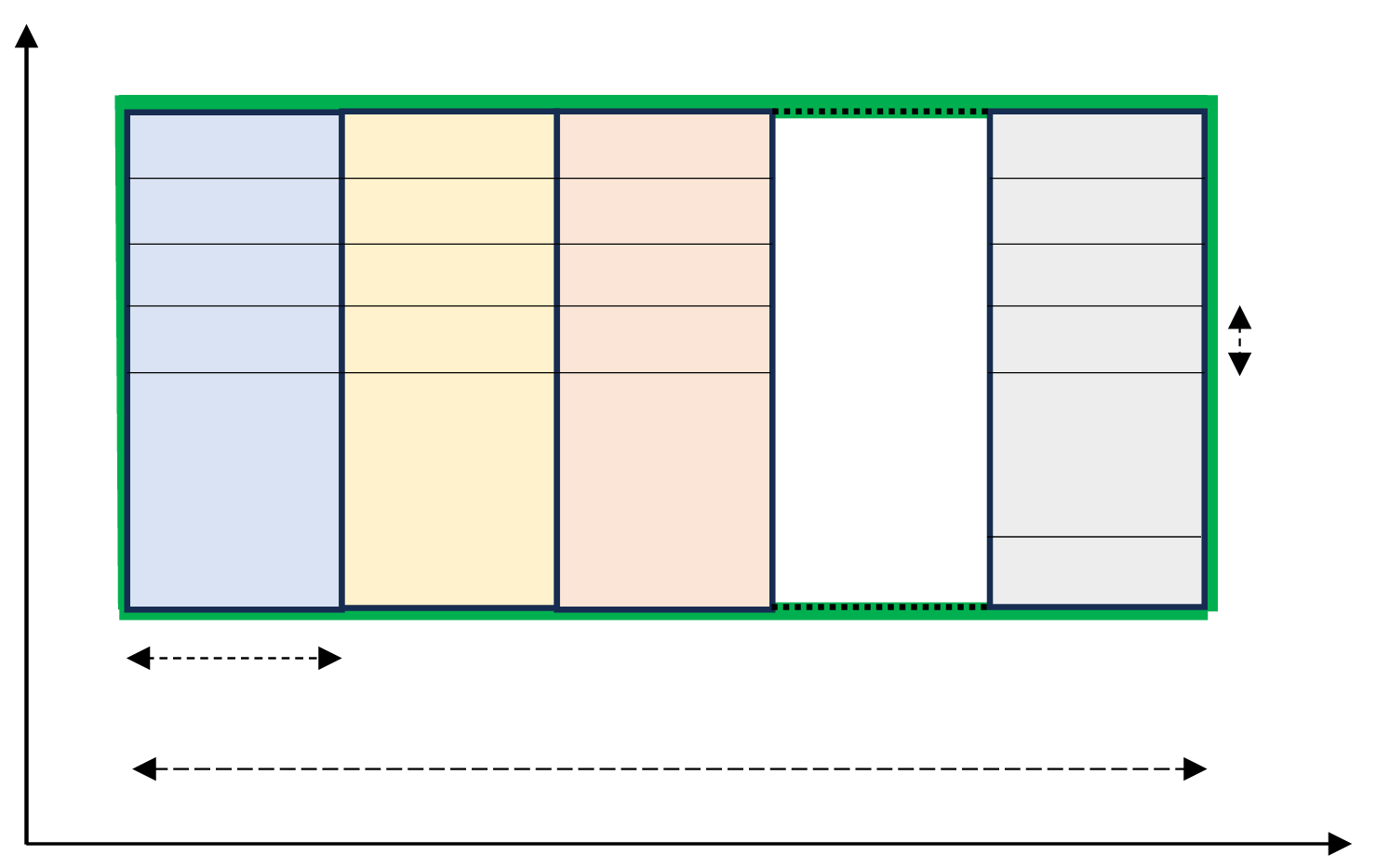}
   \put(13,57){\tiny $\tau =1$}%
   \put(28,57){\tiny $\tau =2$}%
   \put(43,57){\tiny $\tau =3$}%
   \put(62,57){\tiny $\cdots$}%
   \put(75,57){\tiny $\tau =\mathfrak{t}$}%
  \put(14,11){\scriptsize $\mathsf{T}_\mathrm{C}$}%
  \put(16.5,28){\scriptsize $\vdots$}%
  \put(32.5,28){\scriptsize $\vdots$}%
  \put(48,28){\scriptsize $\vdots$}%
  \put(61,39){\scriptsize $\cdots$}%
  \put(80,28){\scriptsize $\vdots$}%
  \put(45,4){\scriptsize $\mathsf{T}_\mathrm{B}$}%
  \put(92,38){\scriptsize $\nu$-th subcarrier}%
   \put(98,2){Time}
  \put(1,62){Frequency}%
   \end{overpic}
\caption{ The channel is approximately time-invariant in each block $\tau$, comprising all subcarriers $S$ and channel coherence time $\mathsf{T}_\mathrm{C}$. The compressed digital combining matrix must be updated at the larger time intervals called the beam coherence time $\mathsf{T}_\mathrm{B}$, which is $\mathfrak{t}$ times larger than  $\mathsf{T}_\mathrm{C}$.}
\label{fig:timeblock}
\vspace{-3mm}
\end{figure}

We consider a wideband mmWave single-user point-to-point MIMO (SU-MIMO) system with $N_\mathrm{t}$ transmit antennas at the base station (BS) and $N_\mathrm{r}$ receive antennas at the UE. {\color{black}As depicted in Fig.~\ref{fig:systemmodel}, we focus on a mobile system in which the scheduled UE moves along a trajectory, while multi-user scenarios will be considered in future work.} The propagation environment is modeled geometrically using  $N_\text{cl}$ scattering clusters \cite{rappaport2015millimeter}. We assume an OFDM signal with $S$ subcarriers and $L$ time-domain taps. The channel matrix undergoes continuous variations over time, which we model piecewise constantly. As depicted in Fig.~\ref{fig:timeblock}, we consider time-domain blocks numbered by $\tau$ that contain all $S$ subcarriers and have a time duration that matches the coherence time $\mathsf{T}_\mathrm{C}$ of the channel \cite{bjornson2017massive}. The frequency domain channel matrix for the $\nu$-th  subcarrier in the $\tau$-th block is denoted as \cite[Ch.~7]{bjornson2024introduction}
\begin{equation}
 \mathbf{H}[\tau,\nu] = \sum_{i=0}^{N_{\text{cl}}} \Big( \underbrace{\sum_{\ell=0}^{L-1} {\alpha_i}[\tau,\ell] e^{-j2\pi \ell v/S}}_{\Bar{\alpha}_i[\tau,\nu]} \Big) {\mathbf{a}_{\mathrm{r}}} ({\phi_{i}^{\mathrm{r}}}[\tau]) \mathbf{a}_{\mathrm{t}}^{\mathrm{T}} ({\phi_{i}^{\mathrm{t}}}[\tau]),   \label{eq:OFDM_ChannelModel}
\end{equation}
for  $\nu = 0,\ldots, S-1$. Here, ${\alpha_i}[\tau,\ell]\thicksim \mathcal{CN}(0, {\beta_{i}}[\tau,\ell])$ is the small-scale fading coefficient of the $\ell$-th time-domain tap and $\beta_{i}[\tau,\ell] \triangleq \mathbb{E}\{ |\alpha_i[\tau,\ell]|^2 \}$ denotes the average power from the $i$-th cluster in the $\ell$-th tap.  $\beta_{i}[\tau,\ell]$ will vary gradually from block $\tau$ to other blocks. 
The vectors $\mathbf{a}_{\mathrm{r}}(\phi_{i}^{\mathrm{r}}[\tau])$ and $\mathbf{a}_{\mathrm{t}} (\phi_{i}^{\mathrm{t}}[\tau])$ are the array response vectors at the UE and BS, respectively. Both the UE and the BS employ horizontal uniform linear arrays (ULA) configuration with antenna spacing $\delta$ so that\footnote{This assumption is made to make the notation tractable, but can be easily extended to uniform planar arrays or other array geometries.}  \cite{bjornson2024introduction}
\begin{align}
\mathbf{a}_{\mathrm{r}}(\phi_{i}^{\mathrm{r}}[\tau]) & = [1, e^{j 2\pi \delta \mathrm{sin}(\phi_{i}^{\mathrm{r}}[\tau]) / \lambda_\mathrm{c}},\ldots, e^{j 2\pi \delta(N_{\text{r}}-1) \mathrm{sin}(\phi_{i}^{\mathrm{r}}[\tau]) / \lambda_\mathrm{c}}]^{\mathrm{T}}, \label{eq:ar}  \\ 
\mathbf{a}_{\mathrm{t}} (\phi_{i}^{\mathrm{t}}[\tau]) & 
=[1, e^{j 2\pi \delta \mathrm{sin}(\phi_{i}^{\mathrm{t}}[\tau])/\lambda_\mathrm{c}},\ldots, e^{j 2 \pi  \delta (N_{\text{t}}-1) \mathrm{sin}(\phi_{i}^{\mathrm{t}}[\tau])/\lambda_\mathrm{c}}]^{\mathrm{T}} ,  \label{eq:at} 
\end{align} 
where $\phi_{i}^{\mathrm{r}}[\tau]$ and $\phi_{i}^{\mathrm{t}}[\tau]$ denotes the azimuth angle of arrival (AoA) and the azimuth angle of departure (AoD) measured from the broadside direction of the respective arrays in the $\tau$-th block, and $\lambda_\mathrm{c}$ is the wavelength at the carrier frequency $f_\mathrm{c}$. The line-of-sight path is denoted by $i = 0$ in \eqref{eq:OFDM_ChannelModel}, where ${\alpha}_0[\tau,0] = \sqrt{\beta_{0}}$, ${\alpha}_0[\tau,\ell] = 0$ for $\ell = 1,\ldots, L-1$, and $\beta_{0}$ describes the large-scale fading. For simplicity, a ULA is considered; however, the proposed methods in this paper are applicable to arbitrary array geometries.

The channel matrix in \eqref{eq:OFDM_ChannelModel} changes continuously over time. The parameters $\Bar{\alpha}_i[\tau,\nu]$ undergoes rapid fluctuations, while the AoA $\phi_{i}^{\mathrm{r}}[\tau]$ and the AoD $\phi_{i}^{\mathrm{t}}[\tau]$, and also $\beta_{i}[\tau,\ell]$, evolve slowly as they are determined by the large-scale geometry. The beam coherence time $\mathsf{T}_\mathrm{B}$ was defined in \cite{khorsandmanesh2024beam} as the duration over which the angular directions remain approximately fixed from a beamforming perspective.
 As indicated in Fig.~\ref{fig:timeblock}, $\mathsf{T}_\mathrm{B}$, is much larger than$\mathsf{T}_\mathrm{C}$ ($\mathfrak{t}=\mathsf{T}_\mathrm{B}/\mathsf{T}_\mathrm{C}$ times larger, shown by the green box). It is seconds instead of milliseconds \cite{khorsandmanesh2024beam}.

\vspace{-1mm}
 \begin{figure*}
        \centering
        \subfloat[Proposed setup]{
   \begin{overpic}[width=1.2\columnwidth]{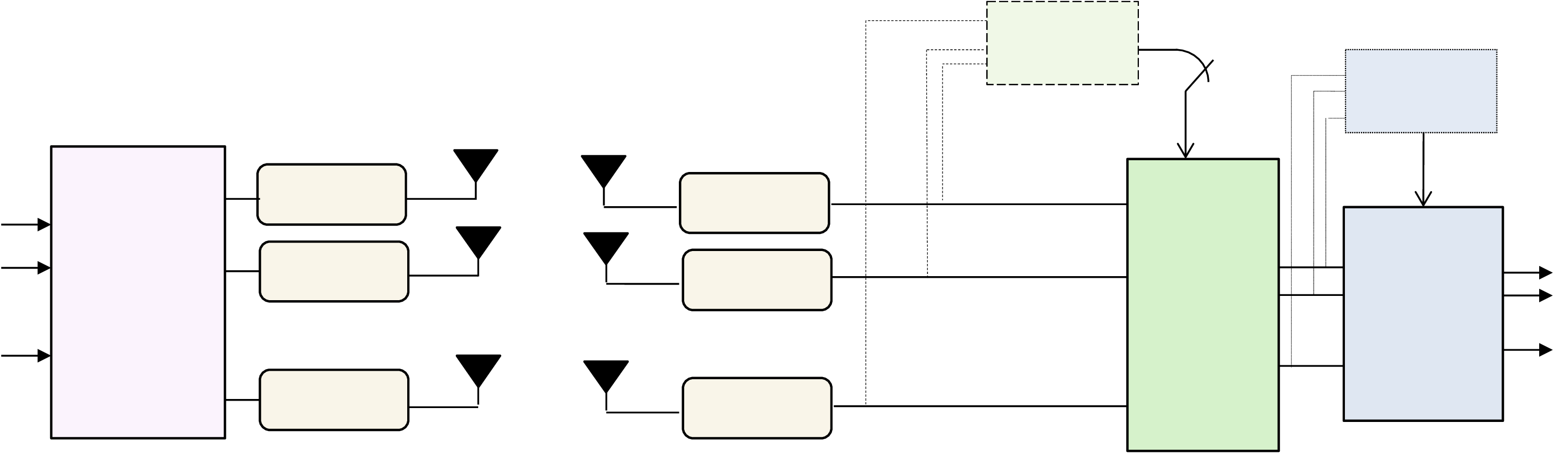}
\put(-1,14.5){\tiny $1$}
\put(-1,12){\tiny $2$}
\put(0.5,7.5){\tiny $\vdots$}
\put(-2.5,6.5){\tiny $N_\mathrm{s}$}
\put(5,15){\tiny  Digital}
\put(5,12.5){\tiny  Precoding}
\put(5,10){\tiny  $\mathbf{F}[\tau,\nu]$}
\put(32,19.5){\tiny $1$}
\put(32,14.5){\tiny $2$}
\put(30,7.5){\tiny $\vdots$}
\put(31,6.5){\tiny $N_\mathrm{t}$}
\put(32,11){\tiny mmWave}
\put(32,9){\tiny Channel}
\put(36.5,19.5){\tiny $1$}
\put(36.5,14.5){\tiny $2$}
\put(39,7.5){\tiny $\vdots$}
\put(36.5,6.5){\tiny $N_\mathrm{r}$}
\put(18,16){\tiny  RF Chain}
\put(18,11){\tiny  RF Chain}
\put(18,3){\tiny  RF Chain}
\put(45,15.5){\tiny  RF Chain}
\put(45,10.5){\tiny  RF Chain}
\put(45,2.5){\tiny  RF Chain}
\put(84.5,12.5){\tiny $1$}
\put(84,10.5){\tiny $2$}
\put(84.5,7){\tiny $\vdots$}
\put(82.5,6.5){\tiny $N_\mathrm{c}$}
\put(65.5,27){\tiny  CSI of }
\put(66,25){\tiny all $N_\mathrm{r}$}
\put(76,28){\tiny Occasionally}
\put(77,26){\tiny updated}
\put(73,16){\tiny  First-Stage}
\put(73,14){\tiny Digital}
\put(73,12){\tiny Combining}
\put(73,10){\tiny (Dimension}
\put(73,8){\tiny Reduction)}
\put(73,6){\tiny $\mathbf{Q}[\nu]$}
\put(88.5,23.5){\tiny  CSI of }
\put(89.5,21.5){\tiny  $N_\mathrm{c}$}
\put(87,12){\tiny Baseband}
\put(87,10){\tiny Digital}
\put(87,8){\tiny Combining}
\put(87,6){\tiny $\mathbf{W}[\tau,\nu]$}
\put(99.5,12){\tiny $1$}
\put(99.5,10){\tiny $2$}
\put(97,7){\tiny $\vdots$}
\put(99,6.5){\tiny $N_\mathrm{s}$}
\end{overpic}
            \label{subfig:A}
        }\quad        \subfloat[Implementation of Beammwave's setup]{
            \includegraphics[width=.29\linewidth]{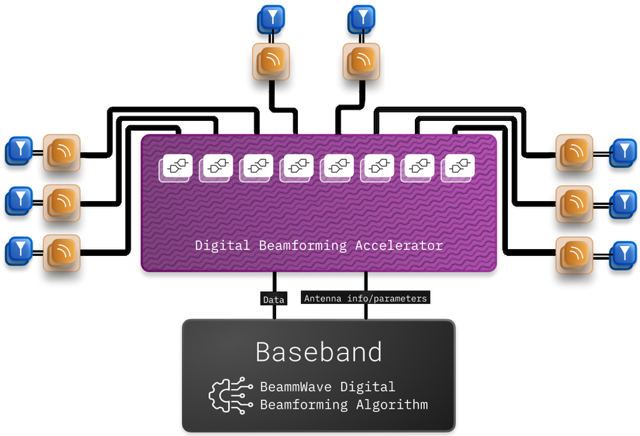}
            \label{subfig:B}
        }
        \caption{{\color{black} Block diagram of a mmWave single-user MIMO system employing fully digital precoding and a two-stage digital combining architecture. a) In the proposed setup, the first-stage combining matrix $\mathbf{Q}[\nu]$ reduces the signal dimension and occasionally accesses CSI across all $N_\mathrm{r}$ antennas. The second-stage combining  $\mathbf{W}[\tau,\nu]$ is updated frequently but has a reduced dimension $N_\mathrm{c} < N_\mathrm{r}$. The beammwave company utilizes the same structure for implementing its UE receiver chips.}}
        \vspace{-5mm}
   \label{fig:hardware}
    \end{figure*}
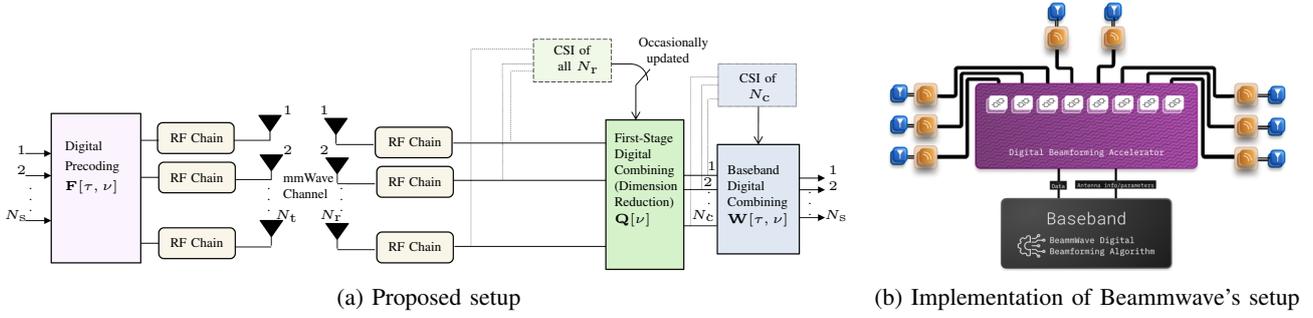

{\color{black}\subsection{Hardware Architecture Properties}}

Even if there has been much prior work on HBF, the long-term goal is to implement DBF in mmWave systems, as its precise control over each antenna element enables fast and flexible adaptation to channel variations \cite{lindoff2021ultimate}. This is possible using state-of-the-art transceiver technology, but the bottleneck is the vast amount of baseband samples that must be processed. A potential solution is to use the receiver architecture shown in Fig.~\ref{subfig:A}, which contains $N_\mathrm{r}$ RF chains and a first-stage digital combining that reduces its dimension to $N_\mathrm{c}$, where $N_\mathrm{s} \le N_\mathrm{c}\le N_\mathrm{r}$, so the rest of the baseband processing (i.e., channel estimation, second-stage combining) can be done with a similar dimensionality as in current HBF methods. Here, $N_\mathrm{s}$ is the number of data streams.
The key to success is that the first-stage combining is implemented efficiently and updated infrequently, so full-dimensional CSI is only required occasionally. {\color{black}Fig.~\ref{subfig:B} shows Beammwave's DBF platform on which the proposed algorithm could be implemented.} 

Based on the beam coherence time concept, it is logical that the first-stage digital combining is updated once per $\mathsf{T}_\mathrm{B}$, while it is fixed within one green block in Fig.~\ref{fig:timeblock}. 
Since the small-scale fading varies more rapidly (i.e., once per coherence time), both the digital precoding $\mathbf{F}[\tau,\nu]$ and second-stage baseband combining $\mathbf{W}[\tau,\nu]$ are updated at this interval for each $\tau$ and $\nu$-th subcarrier.  In contrast, the first-stage digital combining $\mathbf{Q}[\nu]$ remains constant over $\mathfrak{t}$ coherence blocks.

A key difference between the proposed two-stage digital architecture and conventional HBF is that the first-stage digital combining can vary arbitrarily over the subcarriers. By contrast, the analog combining in a hybrid architecture also reduces the dimension but is the same on all subcarriers, since it is implemented using phase shifters.
In the remainder of the paper, we will explore how to operate the considered two-stage digital architecture regarding uplink and downlink channel estimation and what rates are achieved. Note that the BS is assumed to employ DBF without hardware constraints, allowing the focus to remain on the design of the UE's combining. 

\section{Channel Estimation}
\vspace{-1mm}
\label{sec:est}
In this section, we describe how to achieve CSI in the system shown in Fig.~\ref{fig:hardware}. To set the digital precoding $\mathbf{F}[\tau,\nu]$ and the two digital combining $\mathbf{Q}[\nu]$ and $\mathbf{W}[\tau,\nu]$, both the BS and UE need to estimate the channel $\mathbf{H}[\tau,\nu]$. We consider time-division duplexing (TDD) operation, where the channel is first estimated in the uplink and used in the downlink by leveraging channel reciprocity. The first block ($\tau=1$) in the green box in Fig.~\ref{fig:timeblock} is handled differently from the others, as the first-stage combining $\mathbf{Q}[\nu]$ is selected and then remains fixed until the end of the green box. We then estimate the reduced-dimension effective channel $\mathbf{Q}^\mathrm{H}[\nu]\mathbf{H}[\tau,\nu]$ for $\tau >1$. Thus, we consider  $\tau=1$ and $\tau >1$ separately. 
\vspace{-1mm}
\subsection{Channel Estimation in the First Block: $\tau=1$} \label{sec:channelestimate}
In the first block of the $\mathfrak{t}$ that fits in a beam coherence time, the UE transmits pilot sequences on each subcarrier, which the BS uses to estimate the complete channel. Each antenna at the UE transmits an orthonormal pilot sequence of length $\mathsf{t}_{\mathrm{p}} \geq N_\mathrm{r}$. The pilot overhead can be expressed as the ratio $\mathsf{t}_{\mathrm{p}}/\mathsf{t}_\mathrm{c}$, where $\mathsf{t}_\mathrm{c}$ indicates coherence block length in symbols. 
The pilot matrix used by the UE is denoted as $\boldsymbol{\Phi}^{\text{U}}[1,\nu] \in \mathbb{C}^{  N_\mathrm{r} \times  \mathsf{t}_{\mathrm{p}}}$ and has orthonormal rows. The UE transmits $\sqrt{\mathsf{t}_{\mathrm{p}}} \boldsymbol{\Phi}^{\text{U}}[1,\nu]$ to ensure that the total pilot energy is proportional to the pilot length. The received signal at the BS is 
\begin{equation}
\boldsymbol{\mathbf{Y}}^{\text{Pilot,U}}[1,\nu]= \sqrt{P_\mathrm{r}  \mathsf{t}_{\mathrm{p}}} \mathbf{H}^\mathrm{T}[1,\nu] \boldsymbol{\boldsymbol{\Phi}}^{\text{U}}[1,\nu] + \boldsymbol{\mathbf{N}}^{\text{Pilot,U}}[1,\nu], \label{eq:pilotup1} 
\end{equation}
where $P_\mathrm{r} $ is the transmit power used by the UE, normalized by the noise power, and $\boldsymbol{\mathbf{N}}^{\text{Pilot,U}}[1,\nu] \in \mathbb{C}^{{N}_{\mathrm{t}} \times \mathsf{t}_{\mathrm{p}}}$ is the noise matrix at the BS with i.i.d. $\mathcal{CN}(0,1)$-entries. 
Many channel estimators can be developed based on the received pilot signal in \eqref{eq:pilotup1}. In this paper, we adopt the classical ML estimation approach since we consider a mobile UE where prior statistical channel knowledge cannot be obtained. The ML estimate of $\mathbf{H}^\mathrm{T}[1,\nu]$ based on $\boldsymbol{\mathbf{Y}}^{\text{Pilot,U}}[1,\nu]$ is \cite{Kay1993a} 
\begin{equation}  \label{eq:channelMLuplink} \hat{\mathbf{H}}^\mathrm{T}[1,\nu] = \boldsymbol{\mathbf{Y}}^{\text{Pilot,U}}[1,\nu]\boldsymbol{\Phi}^{\text{U}^\dagger}[1,\nu]/ \sqrt{P_\mathrm{r}\mathsf{t}_\mathrm{p}},
\end{equation}
where $(\cdot)^\dagger$   denotes the pseudo-inverse of a matrix. We have ${\boldsymbol{\boldsymbol{\Phi}}}^{\text{U}^\dagger}[1,\nu]= {\boldsymbol{\boldsymbol{\Phi}}}^{\text{U}^\mathrm{H}}[1,\nu] ({\boldsymbol{\boldsymbol{\Phi}}}^{\text{U}}[1,\nu]{\boldsymbol{\boldsymbol{\Phi}}}^{\text{U}^\mathrm{H}}[1,\nu])^{-1} $ in this case.

Now, we shift focus to the downlink data transmission. The received signal on the $\nu$-th subcarrier through $\mathbf{H}[1,\nu]$ for the first block is given as 
\begin{align}
\mathbf{Y}^{\text{D}}[1,\nu] &= \underbrace{\hat{\mathbf{H}}[1,\nu]  \mathbf{F}[1,\nu]}_{\mathbf{B}[1,\nu]}  \mathbf{S}[1,\nu] +  \mathbf{N}^{\text{D}}[1,\nu],  \label{eq:recievedsignal} \vspace{-4mm}
\end{align}
where $\mathbf{S}[1,\nu] \in \mathbb{C} ^{N_\mathrm{s} \times N_\mathrm{d}}$ represents the symbol matrix with independent entries that have unit norms, $N_\mathrm{s} \le \text{rank}(\mathbf{H}[1,\nu])$ is the number of spatially multiplexed data streams transmitted on each subcarrier, and $N_\mathrm{d}$ denotes the number of symbol vectors $\mathbf{s}[1,\nu]\in \mathbb{C} ^{N_\mathrm{s} \times 1}$ (one column of the symbol matrix) that are sent in the downlink. Each entry of $\mathbf{N}^{\text{D}}[1,\nu] \in \mathbb{C}^{ N_\mathrm{r} \times N_\mathrm{d}}$ is i.i.d. $ \mathcal{CN}(0,1)$. The digital precoding $\mathbf{F}[1,\nu] \in \mathbb{C}^{ N_\mathrm{t} \times N_\mathrm{s}}$ is set by the BS utilizing the channel estimates obtained from the uplink in \eqref{eq:channelMLuplink}. Moreover,
$ \Vert \mathbf{F}[1,\nu] \Vert ^2_\mathrm{F}= P_\mathrm{t}$ represents the per-subcarrier transmit
power normalized by
the noise power, where $\Vert \cdot  \Vert_{\mathrm{F}}$ denotes the Frobenius norm. Section~\ref{se:SVD} provides details on how to select the precoding matrix $\mathbf{F}[1,\nu]$.

The notation ${\mathbf{B}}[1,\nu] = \hat{\mathbf{H}}[1,\nu]  \mathbf{F}[1,\nu]$ will be used for the precoded channel. A viable method to provide the UE with CSI is to transmit pilots using the precoder so the UE can estimate ${\mathbf{B}}[1,\nu]$. The downlink pilot matrix is $\boldsymbol{{\Phi}}^{\text{D}}[1,\nu] \in \mathbb{C}^{N_{\mathrm{s}} \times N_{\mathrm{s}}}$. 
By transmitting these pilots over the channel in \eqref{eq:recievedsignal}, the received signal at UE becomes
\begin{equation}
{\mathbf{Y}}^{\text{Pilot,D}}[1,\nu]= \sqrt{ N_{\mathrm{s}}} {\mathbf{B}[1,\nu]}{\boldsymbol{\boldsymbol{\Phi}}}^{\text{D}}[1,\nu] + {\mathbf{N}}^\text{Pilot,D}[1,\nu],
\end{equation} 
where $\boldsymbol{\mathbf{N}}^{\text{Pilot,D}}[1,\nu] \in \mathbb{C}^{{N}_{\mathrm{r}} \times {N}_{\mathrm{s}}}$ is the noise matrix  with i.i.d. $\mathcal{CN}(0,1)$-entries. 
By following the same steps as in the uplink, the estimated precoded channel at $\tau=1$ becomes
\begin{equation}
\hat{\mathbf{B}}[1,\nu] = {\mathbf{Y}}^{\text{Pilot,D}}[1,\nu] {\boldsymbol{\boldsymbol{\Phi}}}^{{\text{D}}^\dagger}[1,\nu] /\sqrt{N_{\mathrm{s}}}.
    \label{eq:estimateB1}
\end{equation}
Now, the UE can select the first-stage digital combining ${\mathbf{Q}}[\nu] \in \mathbb{C}^{N_\mathrm{r} \times N_\mathrm{c}}$ and lower-dimensional second-stage digital combining ${\mathbf{W}}[1,\nu]\in \mathbb{C}^{N_\mathrm{c} \times N_\mathrm{s}}$ based on  $\hat{\mathbf{B}}[1,\nu] $ and ${\mathbf{Q}}^\mathrm{H}[\nu]\hat{\mathbf{B}}[1,\nu]$, respectively. In Section~\ref{se:SVD}, we explain how to select them and why we need the second combining.
\setlength{\parskip}{0pt}
\subsection{Channel Estimation for Blocks  $\tau>1$}
For $ 1<\tau \le \mathfrak{t}$, we need to repeat the steps of uplink channel estimation and estimation of the precoded downlink channel, but now the first-stage combining matrix $\mathbf{Q}[\nu]$ is fixed. We define the effective channel ${\mathbf{G}} [\tau,\nu] = \mathbf{Q} ^\mathrm{H}[\nu]\mathbf{H}[\tau,\nu] \in \mathbb{C}^{ N_\mathrm{c} \times N_\mathrm{t}}$. The UE sends the orthonormal pilot matrix $\grave{\boldsymbol{{\Phi}}}^{\text{U}}[\tau,\nu] \in \mathbb{C}^{{N}_{\mathrm{c}} \times \mathsf{t}_{\mathrm{p}}}$ to estimate ${\mathbf{G}} [\tau,\nu]$. The received signal at the BS is  \vspace{-1mm}
\begin{equation}\grave{\boldsymbol{\mathbf{Y}}}^{\text{Pilot,U}}[\tau,\nu]= \sqrt{P_\mathrm{r}  \mathsf{t}_{\mathrm{p}}} \mathbf{G}^\mathrm{T}[\tau,\nu] \grave{\boldsymbol{\Phi}}^{\text{U}}[\tau,\nu] + \grave{\boldsymbol{\mathbf{N}}}^{\text{Pilot,U}}[\tau,\nu], \label{eq:pilotup} 
\end{equation}
which is similar to \eqref{eq:pilotup1} but contains the lower-dimensional effective channel. Here, $\grave{\boldsymbol{\mathbf{N}}}^{\text{Pilot,U}}[\tau,\nu] \in \mathbb{C}^{{N}_{\mathrm{t}} \times \mathsf{t}_{\mathrm{p}}}$ is the noise matrix with i.i.d. $\mathcal{CN}(0,1)$-entries. The ML estimate of $\mathbf{G}[\tau,\nu]$ is \vspace{-3mm}
\begin{equation}
  \label{eq:estimategt}  \hat{\mathbf{G}}^\mathrm{T}[\tau,\nu] = \grave{\mathbf{Y}}^{\text{Pilot,U}}[\tau, \nu] \grave{\boldsymbol{\Phi}}^{\text{U}^\dagger}[\tau,\nu] /\sqrt{ P_\mathrm{r}\mathsf{t}_{\mathrm{p}}}.
\end{equation}

The BS selects the precoding $\mathbf{F}[\tau,\nu]$ based on $\hat{\mathbf{G}}[\tau,\nu]$ (see Section~\ref{se:SVD} for details). At the UE, the received signal after the first-stage digital combining is
\begin{align}
\grave{\mathbf{Y}}^{\text{D}}[\tau,\nu]  & = \underbrace{\hat{\mathbf{G}}[\tau,\nu]\mathbf{F}[\tau,\nu]}_{{\mathbf{D}} [\tau,\nu]}\mathbf{S}[\tau,\nu] + \underbrace{\mathbf{Q} ^\mathrm{H}[\nu]{\mathbf{N}}^{\text{D}}[\tau,\nu]}_{\grave{\mathbf{N}}^{\text{D}}[\tau,\nu]},  \label{eq:secendeffective}
\end{align} \vspace{-1mm}
where ${\mathbf{D}} [\tau,\nu] = \hat{\mathbf{G}}[\tau,\nu]\mathbf{F}[\tau,\nu]$ is the precoded effective channel. Since the first-stage combining matrix $\mathbf{Q}[\nu]$ only reduces the channel dimension based on the channel geometry but is independent of the current small-scale fading realizations, we also need a second-stage combining matrix ${\mathbf{W}}[\tau,\nu]$ to mitigate interference between the $N_\mathrm{s}$ streams. We need to estimate ${\mathbf{D}} [\tau,\nu]$ to implement that.
\setlength{\parskip}{0pt}
Now, by transmitting the downlink pilot matrix $\grave{\boldsymbol{\Phi}}^{\text{D}}[\tau,\nu] \in \mathbb{C}^{N_{\mathrm{s}} \times N_{\mathrm{s}}  }$, the received signal after first-stage  combining is
\begin{equation} \label{eq:pilot-correlated}
\grave{\mathbf{Y}}^{\text{Pilot,D}}[\tau, \nu]= \sqrt{  N_{\mathrm{s}}} {\mathbf{D}}[\tau,\nu]\grave{\boldsymbol{\Phi}}^{\text{D}}[\tau,\nu] + \grave{\mathbf{N}}^\mathrm{Pilot,D}[\tau,\nu],
\end{equation}
where $\grave{\mathbf{N}}^\mathrm{Pilot,D}[\tau,\nu] \in \mathbb{C}^{N_{\mathrm{c}} \times N_{\mathrm{s}} }$ is colored processed noise with independent columns and the covariance matrix $\mathbf{C}_{\grave{\mathbf{N}}^\mathrm{Pilot,D}} = \mathbb{E} \left[ \mathbf{Q}^\mathrm{H}[\nu]\mathbf{Q}[\nu] \right]$. Since the pilot matrix is multiplied from the right in \eqref{eq:pilot-correlated} and the noise correlation appears from the left, the correlation has no impact on the estimator. The ML estimate of effective downlink channel $\mathbf{D}[\tau,\nu]$ is 
\begin{equation}
\hat{\mathbf{D}}[\tau,\nu] = \frac{ \grave{\mathbf{Y}}^{\text{Pilot,D}}[\tau, \nu] \grave{\boldsymbol{\Phi}}^{\text{D}^{\dagger}}[\tau,\nu]}{\sqrt{ N_{\mathrm{s}}}}. \label{eq:estimateD}  
\end{equation}
All the proposed steps are presented in Algorithm \ref{Alg:ML_Estimation}.

\begin{algorithm}
 [t] 
 \small \caption{Proposed Channel Estimation Approach} \label{Alg:ML_Estimation} \begin{algorithmic}[1] \STATE{\textbf{Input}: Pilot matrices $\boldsymbol{\Phi}^{\text{U}}[1,\nu]$, $\boldsymbol{{\Phi}}^{\text{D}}[1,\nu]$, $\grave{\boldsymbol{\Phi}}^{\text{U}}[\tau,\nu]$,  $\grave{\boldsymbol{{\Phi}}}^{\text{D}}[\tau,\nu]$}   \IF{$\tau = 1$}     \STATE{Estimate uplink channel $    {\mathbf{H}}^\mathrm{T}[1,\nu]$ by utilizing \eqref{eq:channelMLuplink}}    
 \STATE{Set precoding $\mathbf{F}[1,\nu]$ as in \eqref{eq:firstF}}
 \STATE{Estimate downlink channel after precoding $    {\mathbf{B}}[1,\nu]$ as in \eqref{eq:estimateB1}} 
  \STATE{Set compressed first-stage combining  $\mathbf{Q}[\nu]$ as in \eqref{eq:firstQ}}
    \STATE{Set second-stage digital combining  $\mathbf{W}[1,\nu]$ based on lower-dimensional channel $\mathbf{Q}[\nu]\hat{\mathbf{B}}[1,\nu]$ as in \eqref{eq:firstW}}     \ELSIF{$1<\tau \le \mathfrak{t}$}
 \STATE{Define effective channel  ${\mathbf{G}}[\tau,\nu] = {\mathbf{Q}}^\mathrm{H}[\nu]{\mathbf{H}}[\tau,\nu]$} 
\STATE {Estimate uplink effective channel ${\mathbf{G}}^\mathrm{T}[\tau,\nu]$ as in \eqref{eq:estimategt}} \STATE{Update precoding $\mathbf{F}[\tau,\nu]$  as in \eqref{eq:F}}
\STATE {Estimate second effective channel ${\mathbf{D}}[\tau,\nu] $ as in \eqref{eq:estimateD}}
 \STATE{Update second-stage digital combining  $\mathbf{W}[\tau,\nu]$ as in \eqref{eq:w}}
 \ENDIF 
\STATE \textbf{Output}: $\mathbf{F}[\tau,\nu]$, $\mathbf{Q}[\nu]$ and $\mathbf{W}[\tau,\nu]$     
\end{algorithmic}
\end{algorithm}

\section{Downlink Achievable SE}\label{secCapacity}
The proposed channel estimation procedure provides the BS and UE with CSI, which makes the selection of precoding and combining matrices feasible. The available CSI at the UE is crucial in determining the combining and, ultimately, the achievable downlink SE. Therefore, in this section, we will explore the SE that can be achieved with two different levels of CSI availability at the UE: perfect and imperfect CSI.
The achievable SE is upper bounded by the channel capacity \cite{tse2005fundamentals}. In cases where the UE has imperfect CSI, the capacity formula cannot be evaluated directly. However, there are lower bounds that can be utilized to characterize the SE under imperfect CSI. We will apply the \textit{use-and-then-forget} (UatF) technique \cite{bjornson2017massive}. We first consider the ideal scenario where the UE has perfect CSI and describe how to select the precoding and combining matrices in our setup. Then, we present a novel SE expression using the UatF technique.
\subsection{Achievable SE with Perfect CSI at the UE}
We first consider the genie-aided case where the channel ${\mathbf{H}}[\tau,\nu]$ is assumed to be perfectly known to the user. This is a benchmark for the practical estimation method proposed in this paper. The ergodic achievable
SE on the subcarrier $\nu$ is 
\begin{align}
     R[\nu] & =  \mathbb{E} \Bigg[ \log_2 \bigg(\det \Big( \mathbf{I}_{N_\mathrm{s}} +  \big({\mathbf{Q}}[\nu]{\mathbf{W}}[\tau,\nu] \big)^\dagger \mathbf{H}[\tau,\nu] \mathbf{F}[\tau,\nu] \nonumber \\ & \quad \times \mathbf{F}^\mathrm{H}[\tau,\nu]\mathbf{H}^\mathrm{H}[\tau,\nu] \big({\mathbf{Q}}[\nu]{\mathbf{W}}[\tau,\nu] \big) \Big) \bigg) \Bigg],
\end{align}
where the expectation is with respect to the fading process $\{ \mathbf{H}[\tau,\nu] \}$ \cite[Ch.~8]{tse2005fundamentals}. 

The average SE per subcarrier is
\begin{equation}
\text{SE}^{\text{full CSI}} =   \frac{1}{S} \sum_{\nu = 0}^{S-1}  \rho R[\nu],  \label{eq:capacityfullsytem}
\end{equation}
where $\rho = 1- \frac{\mathsf{t}_{\mathrm{p}} + N_\mathrm{s}}{\mathsf{t}_{\mathrm{c}}}$ compensates for the estimation overhead.
\subsection{Precoding and Combining Design} \label{se:SVD}
It remains to select the precoding and combining matrices to maximize the SE, for example, the expression in \eqref{eq:capacityfullsytem}. This is a classical problem with a well-established solution \cite[Ch.~4]{bjornson2024introduction}. 
The optimal approach uses the singular value decomposition (SVD) of each subcarrier’s channel matrix to decouple the MIMO channels into many parallel channels. 
We consider the same approach, but consider the estimated channel matrix. For the $\nu$-th subcarrier and $\tau$-th block, the estimated channel matrix $\hat{\mathbf{H}}[\tau,\nu]$ can be decomposed via the SVD as
\begin{equation}
 \hat{\mathbf{H}}[\tau,\nu] = \mathbf{U}[\tau,\nu]\mathbf{\Lambda}[\tau,\nu]\mathbf{V}^\mathrm{H}[\tau,\nu],
\end{equation}
where ${\mathbf{U}}[\tau,\nu]  \in \mathbb{C}^{N_\mathrm{r} \times N_\mathrm{r}}$ contains the left singular vectors, ${\mathbf{\Lambda}}[\tau,\nu] \in \mathbb{C}^{N_\mathrm{r} \times N_\mathrm{t}}$ is a diagonal matrix containing the singular values in decreasing order on the diagonal, and ${\mathbf{V}}[\tau,\nu]  \in \mathbb{C}^{N_\mathrm{t} \times N_\mathrm{t}}$ contains the right singular vectors. The precoding matrix at the $\nu$-th subcarrier for $\tau=1$ is
\begin{equation}
    \mathbf{F}[1,\nu] =  \mathbf{V}_{(:,N_\mathrm{s})}[1,\nu]\operatorname{diag}(\sqrt{P_{1,\nu}}, \ldots,\sqrt{P_{N_\mathrm{s},\nu}}) , \label{eq:firstF} 
\end{equation}
where $\operatorname{diag}(\sqrt{P_{1,\nu}}, \ldots,\sqrt{P_{N_\mathrm{s},\nu}})$ is an $N_{\text{s}} \times N_{\text{s}}$ diagonal matrix with $P_{i,\nu}$ representing the $i$-th diagonal element. This matrix represents a power allocation over the $i$ streams and follows a water-filling strategy with the
total transmit power constraint $\Sigma^{N_\mathrm{s}}_{i=1} P_{i,\nu} = P_\mathrm{t} $ per subcarrier. Here, $\mathbf{V}_{(:,N_\mathrm{s})}[1,\nu] \in \mathbb{C}^{N_\mathrm{t} \times N_\mathrm{s}}$ contains the first $N_\mathrm{s}$ columns of ${\mathbf{V}}[\tau,\nu]$, corresponding to the dominant $N_\mathrm{s}$ singular values.

To set the first-stage compressed combining matrix, the UE utilizes the SVD of the estimated precoded channel $\hat{\mathbf{B}}[1,\nu] = \mathbf{U}_{\mathrm{B}}[1,\nu]\mathbf{\Lambda}_{\mathrm{B}}[1,\nu]\mathbf{V}_{\mathrm{B}}^\mathrm{H}[1,\nu]$ and pick the first $N_\mathrm{c}$ columns of left singular matrix $\mathbf{U}_{\mathrm{B}}[1,\nu]$ as
\begin{equation}
    {\mathbf{Q}}[\nu] = \mathbf{U}_{\mathrm{B}(:,N_\mathrm{c})}[1,\nu]. \label{eq:firstQ}
\end{equation}

For block $\tau  =1$, we can set the second-stage combining as \begin{align}
  {\mathbf{W}}[1,\nu]  = \mathbf{I}_{N_\mathrm{c} \times N_\mathrm{s} }, \label{eq:firstW}
\end{align}
where $\mathbf{I}_{N_\mathrm{c} \times N_\mathrm{s} }$ denotes a non-square identity-like matrix, due to the SVD-based selection of ${\mathbf{Q}}[\nu]$.

{\color{black}\begin{remark} The channel matrix in \eqref{eq:OFDM_ChannelModel} is 
 \begin{equation}
 \label{eq_ch_compact} 
 \mathbf{H}[\tau,\nu] = \mathbf{A}_\mathrm{r}[\tau] \operatorname{diag}({\boldsymbol{\alpha}[\tau,\nu]}) \mathbf{A}_\mathrm{t}^{\mathrm{T}}[\tau],
 \end{equation}
 where \vspace{-3mm}
 \begin{eqnarray} \label{eq:Ar}
 \mathbf{A}_\mathrm{r}[\tau]  &\triangleq & [\mathbf{a}_\mathrm{r}(\phi_{1}^{\mathrm{r}}[\tau]),\mathbf{a}_\mathrm{r}(\phi_{2}^{\mathrm{r}}[\tau]),\dots,\mathbf{a}_\mathrm{r}(\phi_{N_\text{cl}}^{\mathrm{r}}[\tau])],\\ \label{eq:diagalpha}  \boldsymbol{\alpha}[\tau,\nu] &\triangleq & [\alpha_{1}[\tau,\nu],\alpha_{2}[\tau,\nu] ,\dots,\alpha_{N_\text{cl}}[\tau,\nu]],
 \\ \label{eq:At}
\mathbf{A}_\mathrm{t}[\tau]  &\triangleq & [\mathbf{a}_\mathrm{t}(\phi_{1}^{\mathrm{t}}[\tau]),\mathbf{a}_\mathrm{t}(\phi_{2}^{\mathrm{t}}[\tau]),\dots,\mathbf{a}_\mathrm{t}(\phi_{N_\text{cl}}^{\mathrm{t}}[\tau])],
 \end{eqnarray}
and $\mathrm{diag}({\boldsymbol{\alpha}}[\tau,\nu])$ is the $N_{\text{cl}} \times N_{\text{cl}}$ diagonal matrix with  $\alpha_{i} $ being the $i$-th diagonal element.
The expression in \eqref{eq_ch_compact} acts as an approximate compact SVD when $N_{\text{cl}}$ is small compared to $N_{\text{r}}$, as is the case in mmWave channels. Hence, it is mainly the singular values, determined by ${\boldsymbol{\alpha}[\tau,\nu]}$, that change in each coherence time, and the singular vectors are fixed.
Hence, the first-stage combining based at $\tau=1$ will remain a good choice throughout the beam coherence time.
  
Our numerical results confirm this and the robustness across different scenarios.
\end{remark}}

For $ 1<\tau \le \mathfrak{t}$, the first-stage combining matrix ${\mathbf{Q}}[\nu]$ is fixed. We use the SVD of the estimated effective channel $\hat{\mathbf{G}}[\tau,\nu ] = \mathbf{U}_{\mathrm{G}}[\tau,\nu]\mathbf{\Lambda}_{\mathrm{G}}[\tau,\nu]\mathbf{V}_{\mathrm{G}}^\mathrm{H}[\tau,\nu]$ and the precoding is 
\begin{equation}
    \mathbf{F}[\tau,\nu] =  \mathbf{V}_{\mathrm{G}(:,N_\mathrm{s})}[\tau,\nu] \operatorname{diag}(\sqrt{P_{1,\nu}}, \ldots,\sqrt{P_{N_\mathrm{s},\nu}}),\label{eq:F} 
\end{equation}
using water-filling power allocation.
The second-stage combining is calculated based on estimated precoded effective channel $\hat{\mathbf{D}}[\tau,\nu] = \mathbf{U}_{\mathrm{D}}[\tau,\nu]\mathbf{\Lambda}_{\mathrm{D}}[\tau,\nu]\mathbf{V}_{\mathrm{D}}^\mathrm{H}[\tau,\nu]$ as
\begin{equation}
    {\mathbf{W}}[\tau,\nu] = \mathbf{U}_{\mathrm{D}(:,N_\mathrm{s})}[\tau,\nu]. \label{eq:w}
\end{equation}

\subsection{Achievable SE with Imperfect CSI at the UE}

We will now utilize the UatF approach to characterize the achievable ergodic rate with imperfect CSI. The UE uses the channel estimate to compute the first- and second-stage combining matrices (as described earlier), but 
not when computing the SE. An arbitrary column of the received signal after the second-stage combining can be expressed as 
\begin{align}
& \breve{\mathbf{y}}[\tau,\nu]  = {\mathbf{W}^\mathrm{H}}[\tau,\nu]
{\mathbf{D}}[\tau,\nu]
\mathbf{s}[\tau,\nu] + {\mathbf{W}^\mathrm{H}}[\tau,\nu]   \grave{\mathbf{n}}[\tau,\nu] 
 \nonumber 
\\ &   = \bar{\mathbf{E}}[\nu] \mathbf{s}[\tau,\nu] + (\mathbf{E}[\tau,\nu] - \bar{\mathbf{E}}[\nu]) \mathbf{s}[\tau,\nu]   +  {\mathbf{W}^\mathrm{H}}[\tau,\nu]   \grave{\mathbf{n}}[\tau,\nu],
\label{eq:newnoise}
\end{align}
where $\mathbf{E}[\tau,\nu] = {\mathbf{W}}^\mathrm{H}[\tau,\nu]  {\mathbf{D}}[\tau,\nu] \in \mathbb{C}^{N_\mathrm{r} \times N_\mathrm{r} }$ represents the effective channel after applying the second receive combining matrix and 
$\bar{\mathbf{E}}[\nu]
$ denotes its mean with respect to the fading variations. 
This distinction between the effective channel and its mean is crucial for handling the statistical properties of the channel in subsequent analysis. The term ${\Grave{\mathbf{n}}^{\text{new}}}_d[\tau,\nu] = (\mathbf{E}[\tau,\nu] - \bar{\mathbf{E}}[\nu]) \mathbf{s}[\tau,\nu]       + {\mathbf{W}^\mathrm{H}}[\tau,\nu]   \grave{\mathbf{n}}[\tau,\nu]$ represents spatially colored noise term that 
is uncorrelated with the first term and has the covariance matrix
$\mathbf{C}_{\Grave{\mathbf{n}}^{\text{new}}_d} = \mathbb{E} [ \Grave{\mathbf{n}}^{\text{new}}_d[\tau,\nu]\Grave{\mathbf{n}}^{\text{new}^\mathrm{H}}_d[\tau,\nu] ].$
By interpreting \eqref{eq:newnoise} as a  MIMO channel with the deterministic channel $\bar{\mathbf{E}}[\nu]$ and colored noise, we can now write the average achievable SE of the system as \vspace{-2mm}
\begin{equation}
\text{SE}^{\text{Imperfect CSI
 }} =   \frac{1}{S} \sum_{\nu = 0}^{S-1}  \rho R^{\text{Imperfect CSI}}[\nu] , \label{eq:imperfect}
\end{equation} \vspace{-2mm}
where $\rho = 1- \frac{\mathsf{t}_\mathrm{p} + N_\mathrm{s}}{\mathsf{t}_c}$ and SE at the subcarrier $\nu$ is 
\begin{equation}
R^{\text{Imperfect CSI}}[\nu] =  \log_2 \left( \det \left( \mathbf{I}_{N_\mathrm{s}} + \bar{\mathbf{E}}^\mathrm{H}[\nu] \mathbf{C}_{\Grave{\mathbf{n}}^{\text{new}}_d}^{-1} \bar{\mathbf{E}}[\nu] \right) \right).
\end{equation}
\vspace{-3mm}
\section{Numerical Results}\label{sec5}

In this section, we use Monte Carlo simulations to compare the  SE achieved by the proposed digital precoding/combining architecture with alternative algorithms and HBF.
{\color{black}We simulate a mobile scenario where the UE travels along the linear trajectory, like Fig.~\ref{fig:timeblock}, before reaching the junction.
The BS is positioned at the coordinates (0,0) meters and is equipped with a half-wavelength-spaced ULA comprising $N_{\mathrm{t}} = 64$ antennas. The UE is equipped with a ULA containing $N_{\text{r}} = 16$ antennas aligned with the Y-axis.
It begins its motion at (20,0) meters using a pedestrian speed of $v = 5$ m/s along a vertical line. The movement creates realistic changes in the AoA and AoD, which trigger the need for modifying the precoding and combining matrices. There are $N_{\mathrm{cl}} = 3$ clusters randomly located between the BS and UE. The transmit power is $P_{\mathrm{t}} = 30$ dBm, and the results are presented for $S = 512$ subcarriers. We use the 3GPP pathloss model outlined in \cite[Table 7.4.1-1]{3gpp2018study}, tailored for an urban macrocell environment but without shadow fading. The carrier frequency is $f_\text{c} = 28$ GHz.   We assume $N_\mathrm{s} = 3$, $N_\mathrm{c} = 4$, $\mathsf{t}_{\mathrm{p}}= N_\mathrm{r}$,  and $L=6$. In the next section, we present an average of achievable SE over various random cluster locations and fading realizations.}
\vspace{-2mm}
\subsection{Results and Discussion}
\begin{figure}[!t]
        \centering \includegraphics[width=0.59\columnwidth]{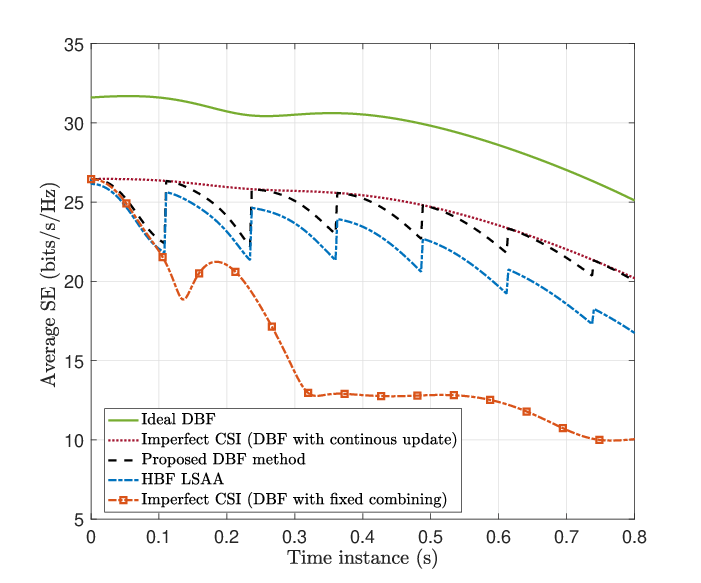}  
  \caption{ The average SE at different time instances when moving along a linear trajectory with speed $\upsilon=5$\,m/s.}
\label{fig:perfect} 
   \vspace{-2mm}
\end{figure}

Fig.~\ref{fig:perfect} shows the average achievable SE of the system over time (in seconds). The top curve is the average SE described in \eqref{eq:capacityfullsytem}, where the UE has perfect CSI and the first-stage digital combining matrix $\mathbf{Q}[\nu]$ is updated every time the channel changes. We call this ``ideal DBF''. The second curve is computed for the proposed method using  \eqref{eq:imperfect}, but updates the first-stage combining every time the channel changes.
The dashed curve represents the proposed method that keeps $\mathbf{Q}[\nu]$ throughout the beam coherence time.
The performance reduces when the channel changes, but the degradation from having a fixed first-stage combining is small, at most 14\%. For comparison, we include SE results using the well-known HBF LSAA approach \cite{sohrabi2017hybrid}, where the first-stage combining is implemented using analog components and has a fixed configuration within the beam coherence time. Our proposed method demonstrates a clear performance advantage over LSAA. The bottommost curve depicts a scenario where both combining matrices, $\mathbf{Q}[\nu]$ and $\mathbf{W}[\tau,\nu]$, remain fixed throughout the simulation. In this case, the mobility significantly impacts the performance, showing substantial degradation over time.

\begin{figure}[!t]
        \centering \includegraphics[width=0.6\columnwidth]{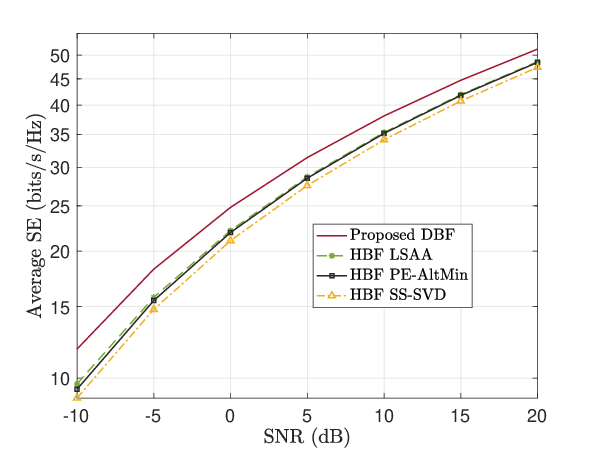}\caption{ The average SE vs. the SNR of the proposed DBF method and different HBF methods with $N_\text{RF}=N_\text{c}$.}
          \vspace{-2mm}
 \label{fig:snr}
\end{figure}

Fig.~\ref{fig:snr} shows the average SE versus the SNR for different combining designs. In this case, we calculate the average SE at the UE coordinates (20,15) meters (i.e., after 3 seconds). We compare our proposed DBF method with LSAA \cite{sohrabi2017hybrid}, the PEAltMin
\cite{yu2016alternating}, and SS-SVD \cite{tsai2018sub}. The results show
that the proposed DBF outperforms other HBF designs at all SNRs. This happens because the proposed digital combining can tune the first-stage combining better to mobile wideband channels.
\vspace{-5mm}
\section{Conclusions}

Hybrid beamforming systems perform poorly in practice due to high heat dissipation and cumbersome mobility support. We have explored an alternative fully digital beamforming architecture with a two-stage combining scheme. The first stage reduces the signal dimensions similarly to the analog part of a hybrid system, but enables higher data rates since the combining can be frequency-selective and contain arbitrary entries.
We demonstrated this by developing an algorithmic single-user MIMO framework with pilot-based channel estimation in both uplink and downlink. If the first-stage combining is updated once per beam coherence time, we can guarantee near-optimal performance even under mobility. The derived SE expression captured imperfect CSI at both the BS and UE, and we show gains over several HBF.

\vspace{-3mm}
\bibliographystyle{IEEEtran}
\bibliography{IEEEabrv,references}

\end{document}